\documentclass[twocolumn,10pt]{IEEEtran}

\pdfoutput=1
\usepackage[cmex10]{amsmath}
\usepackage{amsfonts,amssymb}
\usepackage{multirow} 
\usepackage{makecell} 
\usepackage{booktabs}

\usepackage{verbatim}
\usepackage{url}

\usepackage{cite}
\usepackage{mdframed}

\ifCLASSINFOpdf
  \usepackage[pdftex]{graphicx}
  \graphicspath{{graphics/}}
  \DeclareGraphicsExtensions{.pdf,.jpeg,.png}
\else
  \usepackage[dvips]{graphicx}
  \graphicspath{{graphics/}}
  \DeclareGraphicsExtensions{.eps}
\fi

\usepackage{newfloat}
\DeclareFloatingEnvironment[placement={!t},name=Box]{myBox}

\usepackage{array}

\usepackage{mdwmath}
\usepackage{mdwtab}

\begin{document}
\bibliographystyle{vancouver}
\title{Using Game Theory for Real-Time Behavioural Dynamics in Microscopic Populations with Noisy Signalling}

\author{Adam Noel\IEEEauthorrefmark{1}, Yuting Fang, Nan Yang, Dimitrios Makrakis, and Andrew W. Eckford
	\thanks{\emph{Asterisk indicates corresponding author}.}
	\thanks{\IEEEauthorrefmark{1}A.~Noel is with the School of Engineering, University of Warwick, Coventry, UK (email: adam.noel@warwick.ac.uk).}
	\thanks{Y.~Fang and N.~Yang are with the Research School
		of Engineering, Australian National University, Canberra, ACT, Australia}
	\thanks{D.~Makrakis is with the School of Electrical Engineering and Computer Science, University of Ottawa, Ottawa, ON, Canada.}
	\thanks{A.~W.~Eckford is with the School of Electrical Engineering and Computer Science, York University, Toronto, ON, Canada.}
}


\newcommand{\pop}{\mathcal{A}}
\newcommand{\numPop}{N}
\newcommand{\numPopEst}[1]{\hat{N}_{#1}}
\newcommand{\numCoop}{N_\textrm{c}}
\newcommand{\numGreedy}{N_\textrm{g}}
\newcommand{\numDead}{N_\textrm{d}}

\newcommand{\resource}{R}
\newcommand{\boxWidth}{h}
\newcommand{\energy}[1]{E_{#1}}
\newcommand{\dist}[1]{d_{#1}}
\newcommand{\distMin}{\dist{\textrm{min}}}

\newcommand{\self}{\textrm{s}}
\newcommand{\coop}{\textrm{c}}
\newcommand{\greedy}{\textrm{g}}
\newcommand{\cost}[1]{c_{#1}}
\newcommand{\costCoop}{\cost{\coop}}
\newcommand{\costGreedy}{\cost{\greedy}}
\newcommand{\costSwitch}[1]{s_{#1}}
\newcommand{\costSelf}{\cost{0}}
\newcommand{\reward}[1]{r_{#1}}

\newcommand{\coopFactor}[1]{\gamma_{#1}} 
\newcommand{\coopFactorEst}[1]{\hat{\gamma}_{#1}} 
\newcommand{\interCoop}[1]{\beta_{#1}} 
\newcommand{\selfConversion}[1]{\alpha_{#1}} 

\newcommand{\numRound}{M}


\maketitle

\begin{abstract}
This paper introduces the application of game theory to understand noisy real-time signalling and the resulting behavioural dynamics in microscopic populations such as bacteria and other cells. It presents a bridge between the fields of molecular communication and microscopic game theory. Molecular communication uses conventional communication engineering theory and techniques to study and design systems that use chemical molecules as information carriers. Microscopic game theory models interactions within and between populations of cells and microorganisms. Integrating these two fields provides unique opportunities to understand and control microscopic populations that have imperfect signal propagation. Two examples, namely bacteria quorum sensing and tumour cell signalling, are presented with potential games to demonstrate the application of this approach. Finally, a case study of bacteria resource sharing demonstrates how noisy signalling can alter the distribution of behaviour.
\end{abstract}


\section{Introduction}

For decades, communication engineers have applied mathematics and signal processing to design and understand communication networks. By controlling the end-to-end communication process, engineers have built and continue to design systems that are fast, efficient, and reliable. However, communication system design is not exclusive to engineers. Nature has also evolved many strategies for living things to signal each other and share information.

While it is common knowledge that many species (including ourselves) have natural methods to communicate, some of us may not appreciate the extensiveness and complexity of communication in the microscopic domain, nor the important role it plays, in both our evolution and our everyday health. Signals are being regularly transmitted within and between individual cells and microorganisms. These signals may not be sending packets of data in the conventional communication sense, but nevertheless they enable conventional communication applications such as sensing, coordination, and control. Thus, we can adapt conventional communication engineering theory and techniques to study these signalling mechanisms and understand how they work.

An emerging research field in this direction is \emph{molecular communication}, which considers the use of chemical molecules as information carriers and where traditional communication engineering does not directly apply; see \cite{Farsad2016}. The growth of this field has been primarily driven by two factors. The first is the ubiquitous use of molecules by cells and microorganisms. The second is the incredible potential to use molecules in new devices and networks where traditional communication designs are not suitable, such as for fighting neurological diseases or for sending messages within microfluidic chips. Recent and promising related work includes \cite{Rampioni2018}, where synthetic cells were generated that could communicate with \emph{P. aeruginosa}.

\begin{figure}[!t]
	\centering
	\includegraphics[width=3.5in]{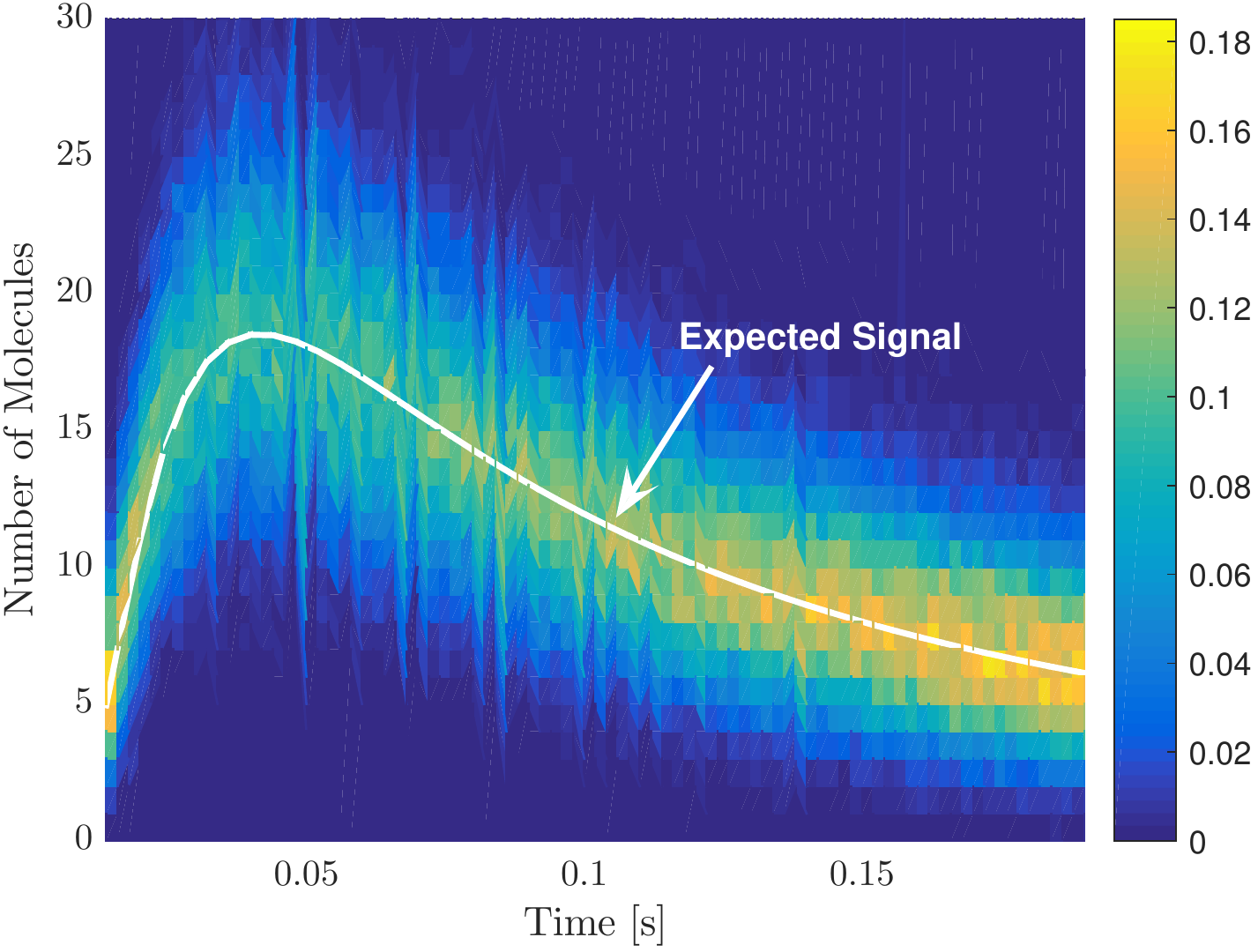}
	\caption{{\bfseries Noisiness of a diffusion wave.} Even in a stable uniform environment without obstacles or chemical reactions, molecular diffusion is a noisy process. The distribution of molecules observed versus time at some distance from an instantaneous point release of molecules is shown. The colour bar on the right is the legend for the distribution values, which add up to 1 for each sampling time. The observed diffusion signal has a variance that is proportional to the time-varying strength of the expected signal (solid white line).}
	\label{fig_diffusion}	
\end{figure}

\begin{figure}[!t]
	\centering
	\includegraphics[width=3.5in]{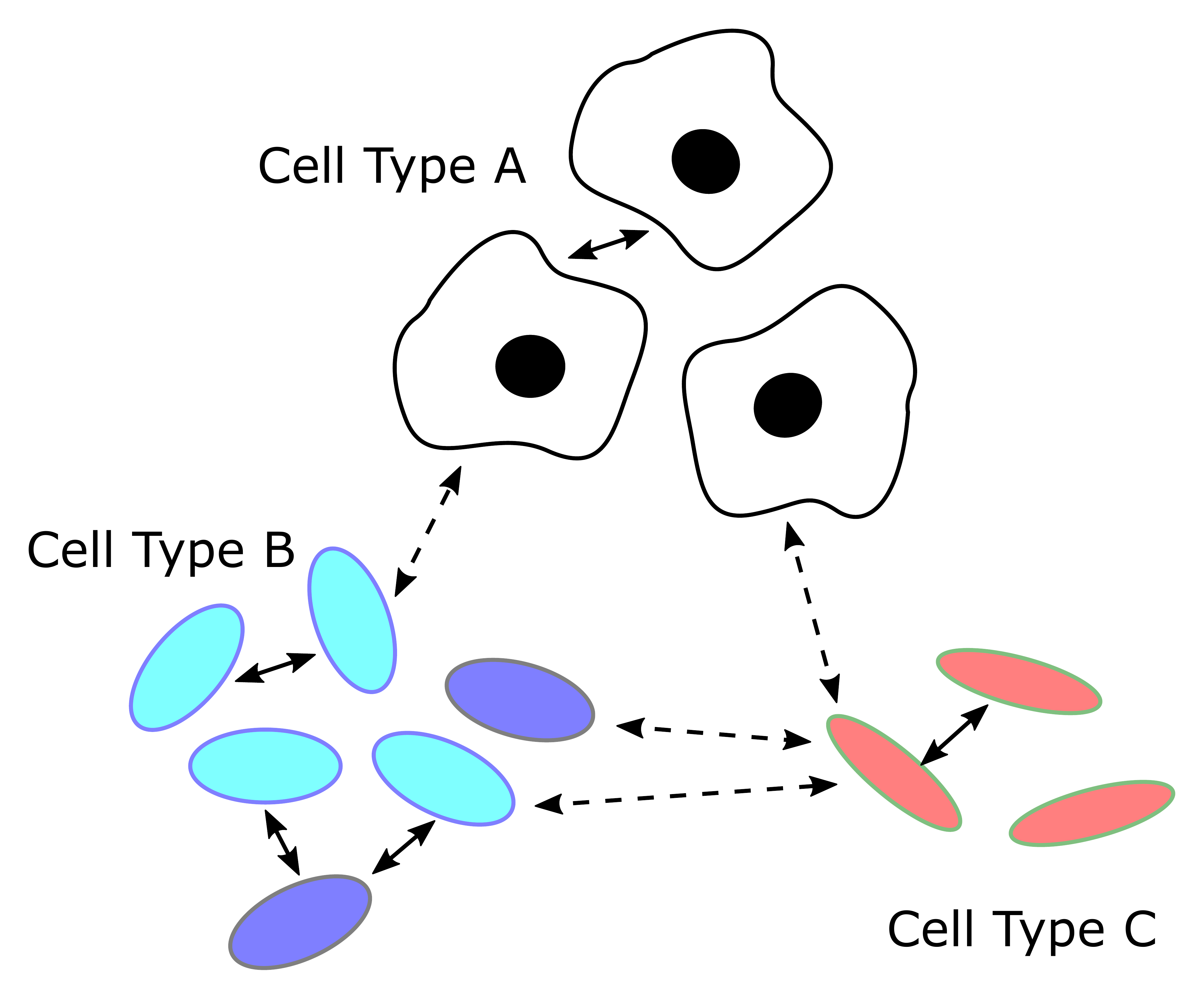}
	\caption{{\bfseries Signalling in a diverse microscopic environment.} Microscopic environments can be home to many different species, including animal cells (Type A) and bacterial cells (Types B and C). Different phenotypes (variations) also express different observable traits (e.g., different shades of cell type B). Cells of the same species commonly communicate with each other (arrows with solid lines), but cross-species communication is also very common, whether intended or not (arrows with dashed lines). Many examples are reviewed in \cite{Atkinson2009}.}
	\label{fig_sketch_cells}	
\end{figure}

An engineer typically expects that a transmitter and a receiver will function as designed. However, unlike modern wireless networks and other communication systems, engineers have less top-down control over systems that include cells and microorganisms. Reasons for this include the limited intelligence of individual cells and the presence of distinct sources of noise. The strength of noise sources, including molecular diffusion and chemical reaction kinetics, are often time-varying and signal-dependent (e.g., see Fig.~\ref{fig_diffusion}). These characteristics lead to the following challenges:
\begin{enumerate}
	\item If we want to communicate with a natural microorganism, such as an individual bacterium, then we are constrained to using (noisy) signalling mechanisms that can propagate in microorganism environments and which they would understand.
	\item Even if signals were correctly received and detected, we may not be able to guarantee that an individual microorganism behaves as intended. Microorganisms do not typically live in isolation but in diverse environments with many species (e.g., see Fig.~\ref{fig_sketch_cells}). Often, these organisms are sharing signals that influence their behaviour, yet they will have imperfect knowledge about each other.
\end{enumerate}

An individual microorganism is not a rational thinker, but it would have evolved to optimize its response to noisy environmental signals. Thus, understanding and controlling microscopic populations requires more than ``simply'' applying communication theory principles. We must also account for the \emph{real-time behavioural dynamics} of the population, i.e., the individuals' responses to repeated interactions, which is the subject of this paper.

The need to predict and control behavioural dynamics suggests the application of \emph{game theory} \cite{Broom2013}. Game theory is a tool for understanding the interactions between rational \emph{players} whose actions (i.e., \emph{strategies}) are influenced by their perceived gains (i.e., \emph{payoffs}). Unlike conventional optimization, game theory models how players adjust their behaviour in response to the behaviour (or \emph{anticipated} behaviour) of other players. For example, in the classical \emph{Prisoner's Dilemma}, two criminals have an incentive to testify against each other, even though the global optimum is for neither of them to testify. Game theoretic models are usually described in terms of strategies and decisions, but they are also applicable (and arguably more so) to microbial populations, even though they are not ``rational'' beings. This is precisely because their behaviours are driven by evolution and responses to external signals; see \cite{Hummert2014}.

This paper serves as an introduction for applying game theory to behavioural dynamics in microscopic environments with noisy signalling. We propose integrating the ideas of game theory for microorganisms with the communication engineering approach from molecular communication. The existing applications of game theory have generally focused on evolution and not accounted for the imperfect propagation of physical signals. Studies of molecular communication have focused on stochastic signal propagation but have not considered behavioural dynamics. This paper seeks to bridge this gap and demonstrate that unique insights and engineering opportunities can result.

We focus on two systems as examples and propose real-time games for these systems. Specifically, we consider quorum sensing by bacteria and signalling by cancer cells. Finally, we use bacteria signalling as a case study and present a corresponding model for the payoffs to individual bacteria. The model extends our preliminary work in \cite{Noel2017c} to show that a higher population density encourages cooperation but that selfish bacteria can also succeed in a dense environment if there are cooperating bacteria to support them. Related work in this area, which did not consider the \emph{control} of behavioural dynamics, includes \cite{Canzian2014,Koca2017,Lindsay2018}. In \cite{Canzian2014}, bacteria decide whether to form links with other bacteria and share resources. In \cite{Koca2017}, two transmitters either compete or cooperate when sending molecules to a common receiver. In \cite{Lindsay2018}, experiments were designed to show that cooperation can be favoured by increasing bacteria population density. Cooperation between bacteria for carrying information is also considered in work including \cite{Unluturk2016a}.

Beyond this paper, our long term objective is to design systems that use chemical signalling, where we can predict and control behaviour between autonomous devices. If we can understand the system as a game, then we can ask how to modify the game in order to achieve a desired result. For example, we could seek how to maintain a healthy system state, how to mitigate disease, or how to efficiently allocate resources for effective signalling. For the bacteria signalling case study considered in this work, we could ask how to facilitate cooperation (e.g., if the bacteria being modelled are healthy for the human body) or how to encourage selfishness (e.g., if the bacteria are infectious).

The remainder of this paper is organized as follows. In Section~\ref{sec_microscopic}, we summarize existing examples of games and game-theoretic analysis in microbial systems. These games highlight the relevance of real-time behavioural dynamics to competition and cooperation. In Section~\ref{sec_properties}, we identify the unique game properties that apply to real-time behavioural dynamics in these environments. In Section~\ref{sec_applications}, we demonstrate the potential to study and control real-time behavioural dynamics through examples of bacteria quorum sensing and cancer cell signalling. We study bacteria signalling as a case study in Section~\ref{sec_case_study}, and conclude in Section~\ref{sec_conclusions}.

\section{Microscopic System Dynamics}
\label{sec_microscopic}

Now we briefly discuss examples of game theoretic applications and related analysis in microscopic systems. These examples demonstrate progress in understanding the complex and dynamic interactions within and between microbial populations, and enable us to draw inspiration to control the real-time dynamics of such populations when they rely on noisy signalling.

Generally, microbial environments are both diverse and dynamic; they are often home to multiple species, including bacteria and animal cells, and their populations can migrate and evolve both spatially and temporally. There are many examples of ``games'' where the players are living cells that compete or cooperate with each other. A common refrain in \cite{Hummert2014} is the importance of spatial heterogeneity; a homogeneous model might predict that only a single behaviour survives, whereas non-uniform spatial distributions can also lead to system stability by providing suitable \emph{local} interactions, i.e., the best behaviour for an individual cell depends on the actions of its immediate neighbours. We now consider some examples.

\subsection{Metabolic Games} There are different metabolic pathways for breaking down sugars into usable energy, including respiration and fermentation. These pathways have different effective rates and different efficiencies (e.g., fermentation is faster but less efficient). We can view a cell's pathway as its strategy, and it is possible for a cell to switch pathways or use multiple pathways simultaneously. \cite{Hummert2014} reviewed games that sought to understand why different pathways have evolved and how different pathways can be maintained within a stable population. For example, it has been shown that fermentation can be favoured in a spatially homogeneous environment; each individual benefits by consuming sugar as fast as it can. However, by accounting for local interactions, it can be shown why and how a mixed population of fermenters and respirators can coexist.

\subsection{Tumor Growth} Recent research has used evolutionary game theory to understand the growth and progression of malignant tumours. \cite{Hummert2014} reviewed games that modelled competition between healthy and tumour cells, and between different types of tumour cells. For example, a tumour can grow by having an equilibrium between cells that are more effective at dividing and cells that are more effective at moving. 

In addition to metabolic games and tumour growth, \cite{Hummert2014} considers how different variations of the same species take turns dominating a population, how different species cooperate to break down resources, and how cells send information with pheromones. The sensitivity to spatial heterogeneity in all of these cases suggests that molecular communication analysis (which models noisy signal propagation between individuals) is relevant for microscopic populations.

\subsection{Quorum Sensing}

A common mechanism for real-time local coordination amongst bacteria is quorum sensing (QS). In QS, each bacterium both releases and detects signalling molecules to estimate the population density. When the density is sufficient, the bacteria initiate collaborative actions, such as biofilm formation. These actions require more effort from each bacterium but can lead to a greater payoff for the community (i.e., a higher chance of survival; see \cite{Bocci2018}). Furthermore, the study of QS has applications beyond bacteria. For example, \cite{Lambert2011} drew analogies between QS and the behaviour of tumours.

Generally, QS mechanisms can be quite complex. \cite{Atkinson2009} described many non-trivial signalling and behavioural dynamics associated with QS, including the use of multiple types of molecules, crosstalk between different species (see Fig.~\ref{fig_sketch_cells}), and eavesdropping by cells that do not release signalling molecules. There are opportunities to model these scenarios as games (as we also discuss later in this paper), and also to draw inspiration from communication engineering concepts such as network security and mitigating interference. Work that has analysed signalling between bacteria as a game includes \cite{Brown2001,Canzian2014}. \cite{Canzian2014} studied the formation of links between pairs of bacteria as a repeated game and whether a colony of connected bacteria could form. \cite{Brown2001} presented a cooperation game that accounts for costs associated with cooperation and signalling molecule generation. Existing analysis does not tend to model physical molecular signals and their stochastic signalling dynamics, although they have been identified as important factors to drive heterogeneity in microbial infections; see \cite{Davis2019}.

\section{Properties of Real-Time Microscopic Games}
\label{sec_properties}

By focusing on the \emph{control} of \emph{real-time} behaviour under \emph{noisy} signalling, our proposed game theoretic approach has features that are distinct from existing analysis of microscopic systems. Here, we highlight both the unique properties of the game theoretic components and the differences in how solutions could be obtained.

\subsection{Players}

Microorganisms qualify as players, even though they are not rational decision-makers. While many games in biology are modelled at the population level, microscopic games that account for local interactions require that the game be modelled at the level of individual cells. This was also needed for the spatially heterogeneous games reviewed in \cite{Hummert2014}. Furthermore, due to the limited intelligence of the individual microorganisms and the impact of information uncertainty due to noisy signalling via stochastic reaction-diffusion, we are interested in games with players that have \emph{imperfect real-time} information about each other. For example, a microorganism will most likely not know the precise number of players nor the actual behaviour of each player (information via cell signalling may be limited to only a few bits; see \cite{Ruiz2018}). This is especially true when the population changes, e.g., players enter or leave via motility, cell division, and death. We also seek to manipulate the propagation of the molecular communication signals so that we can tune the perceived information and control the players.

\subsection{Strategies}

In biological systems, individual players are typically treated in aggregate and one describes the distribution of strategies in a population, e.g., what fraction is cooperating and what fraction is cheating. For microscopic games, we should also consider the spatial distribution of strategies, as emphasized in \cite{Hummert2014}. Additionally, to be \emph{real-time}, we seek to model the dynamics of \emph{individual} behaviour, where a player might change its behaviour due to its ongoing (but noisy) assessment of the environment. Individual dynamics are common in general game theory (as \emph{repeated games}; see \cite{Maschler2013}). However, microscopic game theory problems usually assume that a player's behaviour is fixed and variations are only achieved in future generations via mutations. Part of the novelty of our approach is having a game where a microscopic player's strategy can change \emph{within its lifetime}.

\subsection{Payoffs}

As in existing microscopic games (and in games more generally), the payoffs in our approach are the net rewards that players receive as the outcome of the game, as a function of the strategies of all of the players. Whereas a player's strategy depends on \emph{perceived} information about the system, which can be incorrect, the payoffs depend on the \emph{actual} current system state. Nevertheless, to be relevant to our approach, a suitable payoff model should include the following:
\begin{enumerate}
	\item Accommodate the spatial distribution of players. For example, the reward for a player that is adjacent to a cheater may be less than that for a player that is surrounded by cooperators.
    \item Vary with time, both to model the temporal behaviour of the players, and also to account for a dynamic environment. For example, the players may consume a resource that depletes and replenishes over time.
    \item Impose a cost on a player \emph{changing} its behaviour. Obviously, different behaviours should have different costs (e.g., cooperation is generally modelled to be more resource intensive than cheating). However, it is also reasonable for a player to spend resources in order to switch from one behaviour to another. For microorganisms, this could be represented as the time and energy needed to alter gene expression.
    \item Be tunable. We seek to control the environment by altering the cost or the reward for particular behaviours. For example, we might seek to make cooperation more expensive in order to prevent the formation of a biofilm.
\end{enumerate}

\subsection{Solutions}

Most biological games, including microscopic games, are studied using evolutionary game theory; see \cite{Broom2013, Hummert2014, Maschler2013}. This means that the solution of interest is the Evolutionarily Stable Strategy (ESS), which is a distribution of strategies that remains stable over generations of players. This framework is consistent with a model where a player's behaviour is fixed over its lifetime.

For us to consider real-time games between dynamic microscopic players, the similar but distinct \emph{Nash Equilibrium} (NE) framework is more appropriate. When a NE is achieved, no player can benefit by changing its strategy (unless, of course, the game itself changes). Thus, from the context of solutions, we are interested in how we could guide a microbial population towards a particular NE or how we could convert a desired system state into a NE. For example, if the conditions leading to the formation of a tumour was a NE, then we might seek to prevent this NE by making the requisite cooperation between cancer cells a non-equilibrium state.

\section{Sample Applications of Game Theory and Molecular Communication to Microorganisms}
\label{sec_applications}

In this section, we consider two practical systems where we seek to integrate game theory and molecular communication to control behavioural dynamics in microscopic populations. For each system, we describe example games with both one and multiple species.
The first system considers quorum sensing to achieve cooperation within a bacterial population. We completed a preliminary study of this scenario with a simple resource sharing game in \cite{Noel2017c}, and we extend this model as a case study in Section~\ref{sec_case_study}. The second system considers signals from tumour cells and their interactions with healthy tissue and the immune system. The two systems demonstrate the breadth of opportunities for integrating stochastic signal propagation with game theory. Game theory enables us to model complex interdependent behaviour, and molecular communication analysis enables us to describe the imperfect local information due to stochastic signal propagation.

\subsection{Application 1: Bacteria Signalling}

We are particularly inspired by QS as an implementation for communication between bacteria and are interested in its influence on real-time behaviour. For the first example, we consider a resource sharing game where the players are all members of the same QS population, and then an eavesdropping game with multiple species where only one population uses QS signals. Study of the corresponding models might lead to improved strategies for combating antibiotic resistance or improving the health of essential bacterial communities.

\subsubsection{Resource Sharing Game}

Consider a resource sharing game where bacteria consume a common resource (e.g., food) and they could work together to access the resource. For example, the bacteria could cooperate to coordinate an attack on a larger organism or to optimize nutrient extraction via cross-feeding (see \cite{Hummert2014}). In QS, each bacterium estimates the size of the population. We are interested in how the uncertainty in the population (both size and state) affects the dynamics of the population. If we assume that all bacteria behave in their own interest, then any individual bacterium would only commit the additional resources necessary for cooperation if it would benefit from doing so, or if it ``\emph{believes}'' that it would benefit.

We considered a very simple model of this game in \cite{Noel2017c}, which included only some of the distinctive game properties identified in this paper, and extend the model as a case study in Section~\ref{sec_case_study}. The results in \cite{Noel2017c} suggested that uncertainty in the size and behaviour of the rest of the population can overcome a lack of explicit coordination and lead to cooperation. In Section~\ref{sec_case_study}, we add features to make the model more practical. In particular, the payoffs depend on the local number of bacteria and their behaviour, less information is available for bacteria to infer the state of the population, and conditions are established for bacteria to win or lose (i.e., succeed or die).

In other variations of this game, some bacteria might not fully participate in the resource sharing process, either by never releasing QS molecules or never cooperating. These actions make them ``free-riders'' of any benefit from the cooperating population. This has been observed experimentally in mutants with no QS mechanism in \cite{Popat2012}, and is particularly beneficial to a free-rider if the cost to transmit the QS signal or the cost to cooperate is expensive. A bacterium can still benefit from cooperation if there are still enough bacteria cooperating in the quorum. However, the population estimation will be perturbed and if there are too many free-riders then the bacteria that are still signalling are wasting energy. Open questions include (1) whether game theory can enable us to predict a stable number of free-riders, and (2) whether mechanisms exist to prevent too many bacteria from free-riding.

\subsubsection{Inter-Species Eavesdropping Game}

As we noted previously, QS can include signalling between different species and enable one population to eavesdrop on another. For example, \cite{Atkinson2009} discussed how bacteria such as \emph{E. coli} and even animals such as \emph{C. elegans} do not generate QS signals but can intercept them from other species. Through ``silent'' observation, \emph{E. coli} can mount ``stealth attacks'' on hosts and \emph{C. elegans} can be both attracted to food sources and repelled by pathogens. With these examples in mind, we can consider a game where the QS bacteria have to also decide the strength of their QS signals. A stronger signal can make achieving cooperation more reliable but at the cost of making detection by other populations easier. The members of the eavesdropping population have to decide whether the noisy QS signal is sufficient to take action against the QS bacteria population. Parameters of interest include the size and proximity of these populations, as these would influence the reliability of the signals as well as the payoffs. We can consider whether the bacteria could be manipulated to make them easier to detect and whether the eavesdropping population can adjust its sensitivity to the QS signal to avoid taking action and wasting energy when it is unnecessary.

\subsection{Application 2: Tumor Cell Signalling}

Our next application considers a more diverse environment that includes cancer cells, healthy cells, and immune system cells. Cancerous tumours are groups of cells that undergo abnormal growth and can invade surrounding tissue. They can eventually metastasise and spread throughout the body, at which stage they are very difficult to treat. Thus, we are particularly interested in the formation and behaviour of premalignant tumours. For this application, we consider a diffusion control game that is played by the tumour cell population, and then a more complex game where we add the immune system cells as players. The proposed games use signalling and decision-making at the cellular level to gain insight into cancer development and treatment.

\subsubsection{Diffusion Control Game}

\begin{figure}[!t]
	\centering
	\includegraphics[width=3.5in]{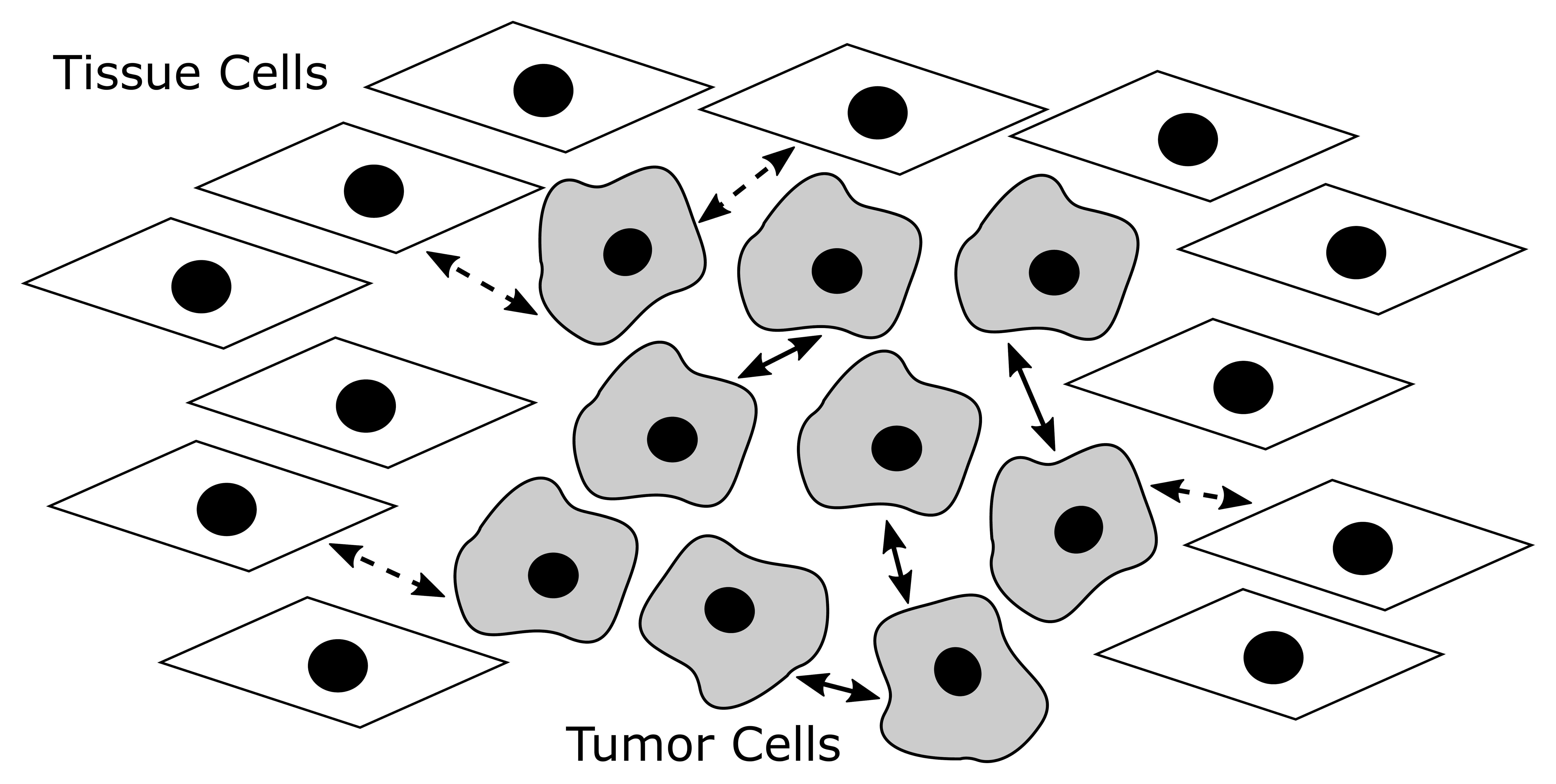}
	\caption{{\bfseries Signalling by tumour cells.} Cancer tumour cells (shaded) are capable of manipulating nearby healthy tissue cells (white) to produce infrastructure that protects the tumour and shields it from detection by the immune system. The signalling by the tumour to the tissue (represented by arrows with dashed lines) results in a environment that has similarities to that achieved by bacteria communities that create biofilms; see \cite{Lambert2011}.}
	\label{fig_sketch_tumour}	
\end{figure}

Tumours are more than just cells that grow and divide without restraint; they are complex communities that signal among themselves and with the surrounding environment. As we previously noted, \cite{Lambert2011} observed that evolutionary strategies used by bacteria can also be identified within tumours. For example, when bacteria create a biofilm, they increase their resistance to external threats such as antibiotics but this also reduces the intake of nutrients and oxygen. Tumour cells undergo a similar trade-off when they stimulate surrounding cells to both produce more extracellular matrix and increase their metabolic rates, which simultaneously reduces access of molecules to the tumour via diffusion (which reduces the ability of the immune system to identify the cancerous cells via antigens) while maintaining energy needs (see Fig.~\ref{fig_sketch_tumour}).

Along this direction, we could model tumour growth as a diffusion control game where tumour cells choose whether to stimulate the surrounding tissue, i.e., whether to reduce the diffusion rate (which reduces the ambient nutrient levels but decreases the chance of detection by the immune system). The player model could include a tumour cell's uncertainty about its population size and its own location in the tumour relative to the surrounding tissue. These parameters could be estimated via molecular communication, i.e., inferred from the concentrations of molecules in the vicinity of the player. The payoff model could include the cost of signalling to the tissue while showing the expected trade-off between detectability and access to nutrients. This game could help us understand and mitigate the conditions where healthy tissue supporting the tumour is an attainable NE, for example by increasing the cost needed to reliably communicate with the healthy tissue.

\subsubsection{Competition with Immune System}

We can extend the diffusion control game by adding the immune system, which provides both adaptive and innate protection against external threats; see \cite{DeVisser2006}. An initial game theoretic model in this direction could consider the energy costs associated with building an adaptive immune response versus the response's capacity to detect and fight cancer before a tumour can metastasise. The dynamics of this model would include (1) determining the number of adaptive cell players to respond to the detection of a tumour, (2) the strategy of each adaptive immune cell to identify and respond to the tumour, and (3) the tumour cells' efforts to protect themselves and whether they are able to detect the immune system's behaviour. Each of these components would rely on noisy observations of propagating signals, e.g., the probability of tumour detection would rely on how easily antigens can reach and identify the tumour. Another work that considered this problem but not within the context of a game is \cite{Hsu2017}.

There is significant potential to explore this model, as\cite{DeVisser2006} reviewed epidemiological studies demonstrating that patients with compromised adaptive immunity can be at a \emph{reduced} risk for some types of cancers, and environmental conditions can actually prompt innate immune cells to \emph{promote} tumour growth by suppressing an adaptive response. This suggests that individual immune cells can indeed be modelled as players that could be manipulated to fight or support a tumour. Furthermore, the model could be integrated with clinical tools such as immunotherapy, where the adaptive immune system is modified to improve immunity against a particular target; see \cite{DOnofrio2005}.

\section{Case Study: Bacteria Resource Sharing}
\label{sec_case_study}

We complete this paper with a case study of bacterial signalling. The case study is a more practical extension of our preliminary model in \cite{Noel2017c}. In particular, the entire game is modelled at the level of individual bacteria (such that payoffs are heterogeneous, even when bacteria have the same behaviour), bacteria use a noisy estimate of their current payoff to infer the state of the population, and bacteria can win or lose (i.e., succeed or die) as the game progresses. It is still a simple model, as we make many simplifying assumptions about the bacteria and we use rudimentary means to account for the signal propagation. Nevertheless, the model is sufficient to obtain interesting results that are consistent with intuition.

In this section, we summarize the system model and corresponding game, present and analyse simulation results, and provide comments on future directions for the model. For clarity of presentation, the full technical details of the model are described in the Appendix.

\subsection{Model and Game Summary}

We consider a static population of bacteria players. Every bacterium has an energy level; it gains energy from the environment and spends energy to operate. A player can die if it runs out of energy or succeed if it collects a sufficient amount. By ``succeed,'' we mean that a player has sufficient energy to survive or proliferate and no longer participates as a formal player in the game.

We model the inter-dependence of players via the energy collection process and in consideration of their proximity because cell fitness has been shown to depend on cell density; see \cite{Lindsay2018}. The strategy of every player is to be either greedy or cooperative. Greedy players require less energy than cooperators to operate. Generally, it is possible for proximity to be either positive or negative for either cooperating or greedy players; see \cite{Lindsay2018} for examples and experiments demonstrating each case. In this case study, we allow cooperative players to benefit by being close to each other (because they work together to improve resource access) whereas greedy players benefit by being separated (because they compete for resource access).

The heterogeneity in the energy collection model also drives asymmetry in the distribution of behaviour. The players estimate the amount of collected energy; this estimate can also be corrupted by additive white Gaussian noise (AWGN). The estimate is used to infer the size and behaviour of the total population by assuming that all other players are nearby and have homogeneous behaviour. From these inferences, a player compares its potential energy income from either behaviour and switches if it is both beneficial to do so and if the player can afford a ``switching fee''.

The game is played in multiple rounds. In every round, each player has its energy updated and it estimates the population state to determine its behaviour in the following round. The rounds continue until all the players are starved or successful, or until a maximum number of rounds has occurred.

\subsection{Results and Discussion}

We now consider simulations of the proposed game. We simulate the game where we randomly place the bacteria over a square region and they remain fixed for the rounds of that game. The system model parameters are configured as shown in Table~\ref{table_sim_parameters} in the Appendix. Each game begins with 50 cooperative and 50 greedy bacteria, and every player re-estimates the state of the population (i.e., size and behaviour) in every round.

Since we are interested in the behavioural dynamics, we seek to characterize the changes in behaviour as the game progresses, the eventual number of starved and successful players, and which behaviour led to starvation or success. Thus, for each configuration we plot the number of playing, starved, and successful players of each behaviour as a function of the game round. Games are repeated (and locations re-generated) 100 times and all curves show the state of the players averaged over all games. Error bars show one standard deviation and are omitted when the standard deviation is less than 0.5 players.

In this preliminary analysis, we focus on three features. First, we set the average population density by controlling the size of the square region over which the bacteria are distributed. Second, we set whether the switching costs are sufficiently low to enable switching (if not, then the system progresses without an actual game being played). Finally, we set whether the collected energy is observed perfectly or with AWGN.

\begin{figure}[!t]
	\centering
	\includegraphics[width=3.5in]{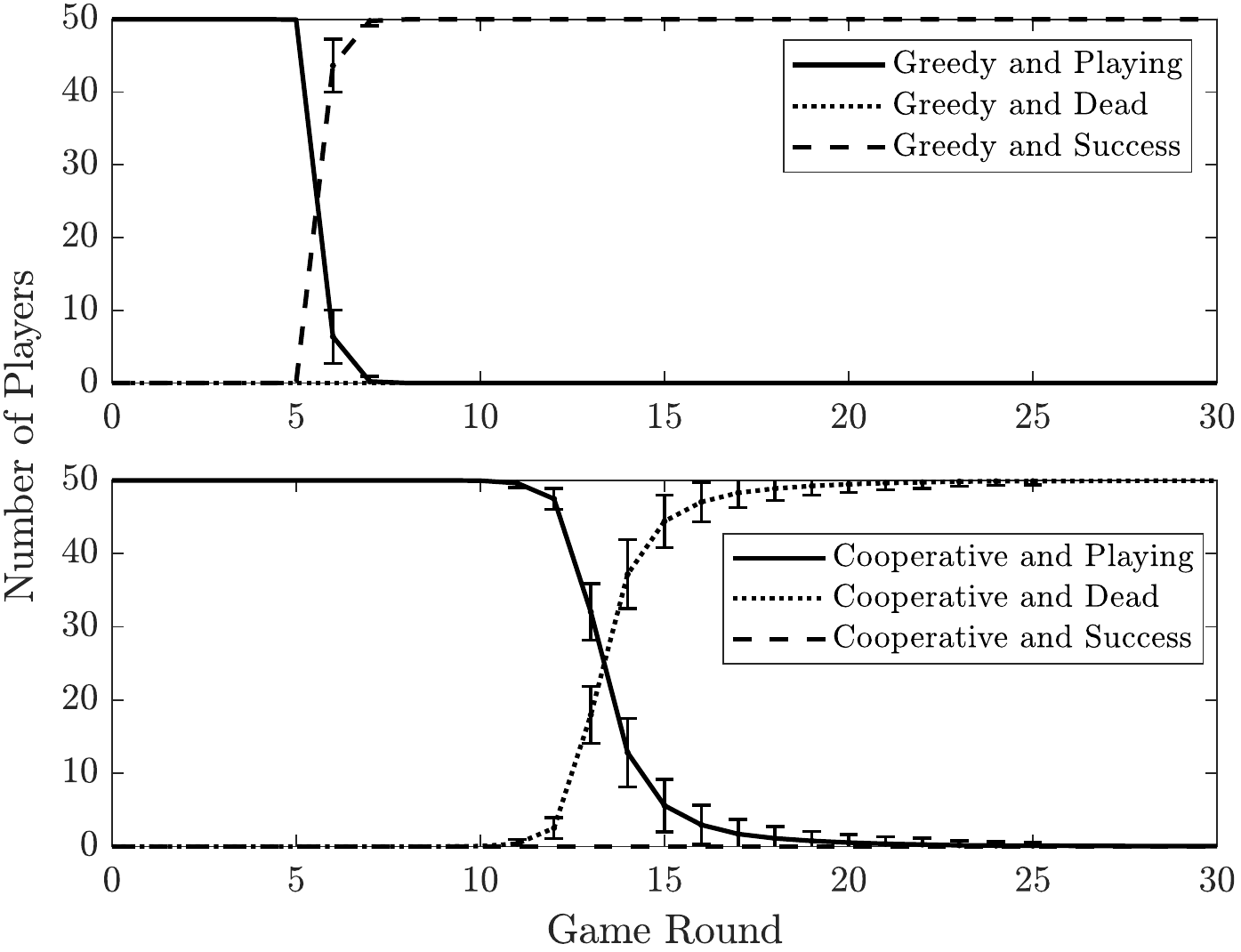}
	\caption{\textbf{Behaviour with low density and no switching.} The cooperative players all die and the greedy players all succeed.}
	\label{fig_bacteria_game_low_den_high_switch}
\end{figure}

\begin{figure}[!t]
	\centering
	\includegraphics[width=3.5in]{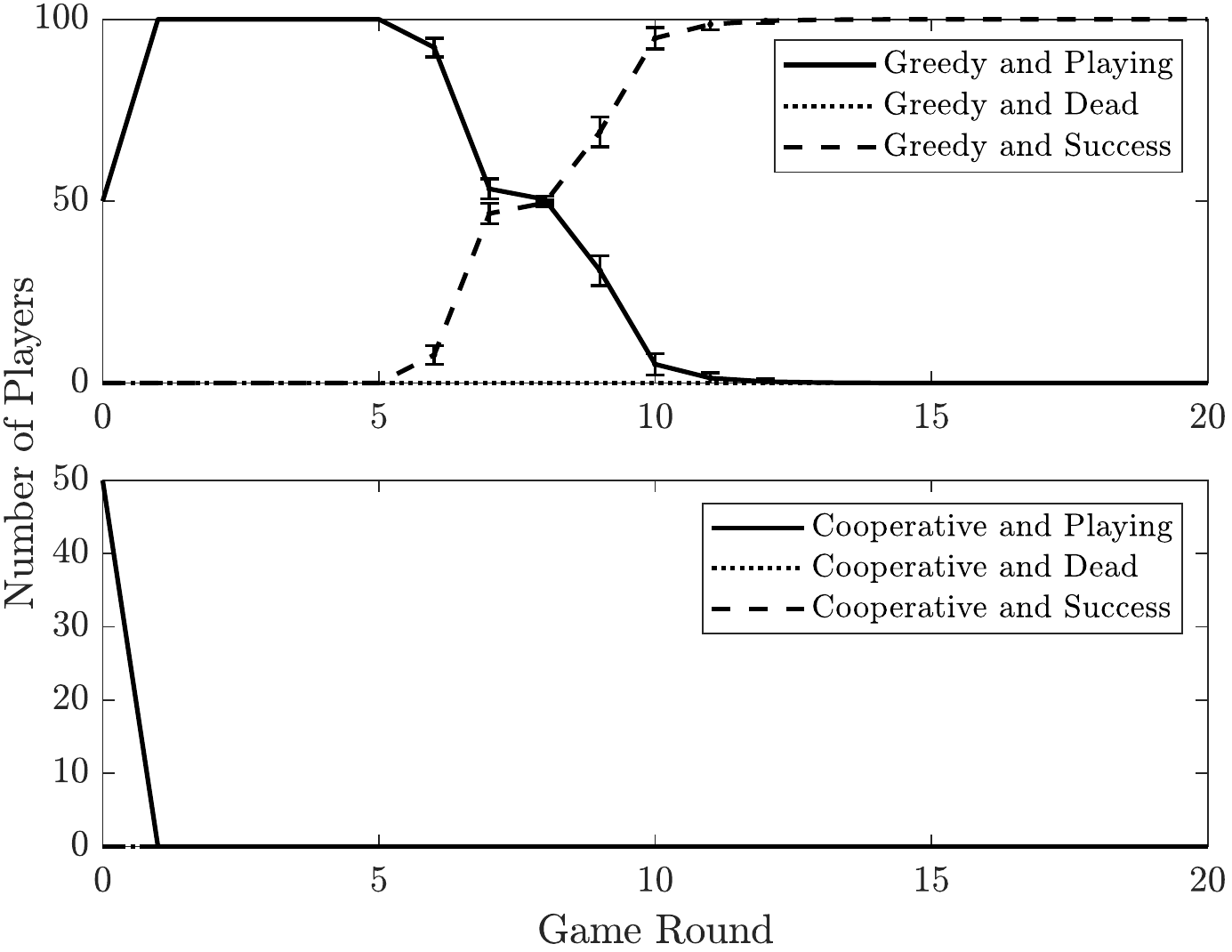}
	\caption{\textbf{Game with low density.} All players become greedy and succeed.}
	\label{fig_bacteria_game_low_den_low_switch}
\end{figure}

\begin{figure}[!t]
	\centering
	\includegraphics[width=3.5in]{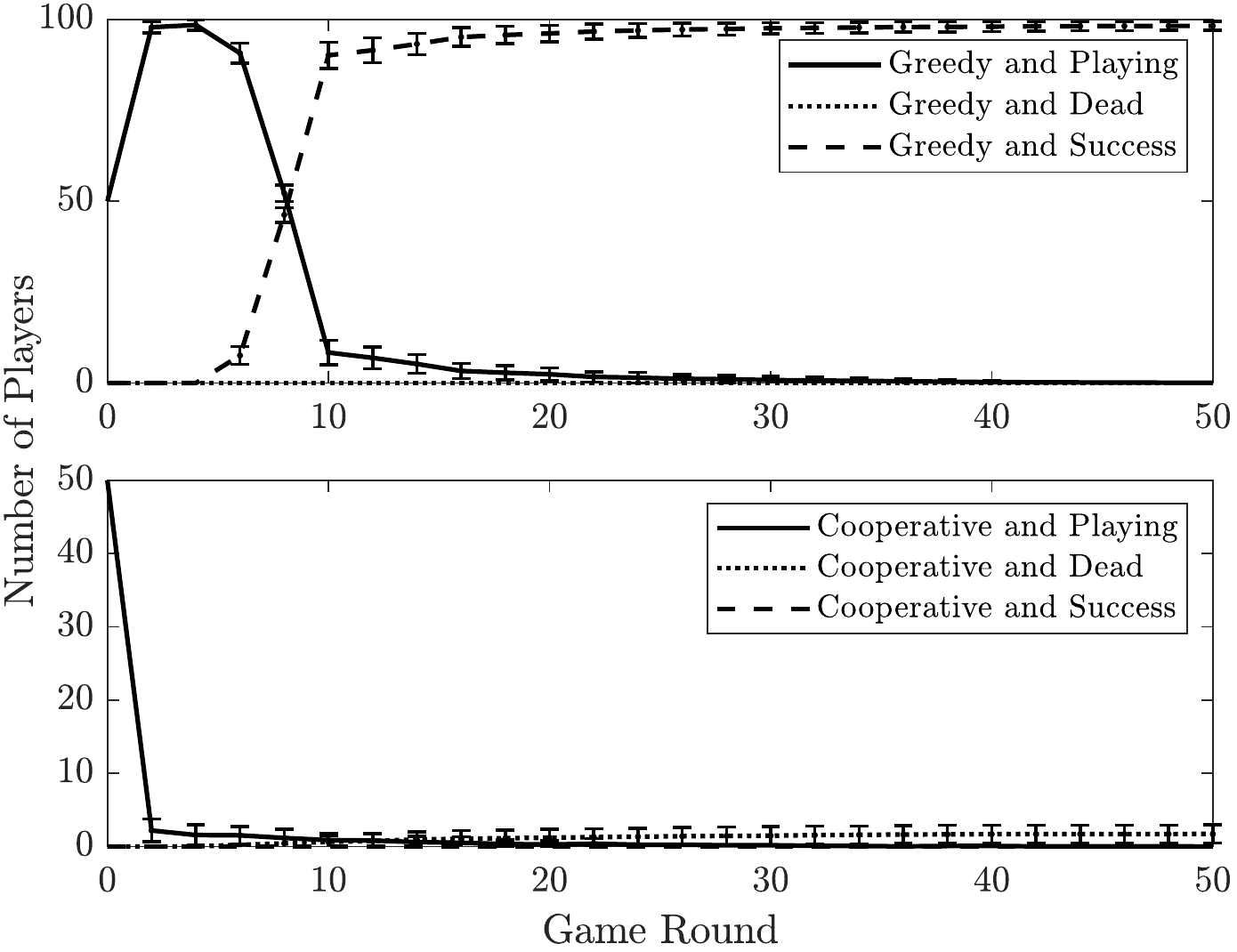}
	\caption{\textbf{Game with low density and estimation with AWGN.} Most players become greedy and eventually succeed; players that stay cooperative die.}
	\label{fig_bacteria_game_low_den_low_switch_noisy}
\end{figure}

We consider a low average population density in Figs.~\ref{fig_bacteria_game_low_den_high_switch}, \ref{fig_bacteria_game_low_den_low_switch}, and \ref{fig_bacteria_game_low_den_low_switch_noisy}, where the players are distributed over a square of length $100\,\mu$m. Based on our model, we expect that greedy behaviour would be more successful. This is confirmed in Fig.~\ref{fig_bacteria_game_low_den_high_switch}, where the players are prevented from switching their behaviours. The greedy players all succeed within 8 rounds, whereas the cooperative players are all dead within 25 rounds. When the switching costs are lowered, as considered in Fig.~\ref{fig_bacteria_game_low_den_low_switch}, all of the players switch to greedy and eventually succeed. However, when we then introduce AWGN to the population estimation in Fig.~\ref{fig_bacteria_game_low_den_low_switch_noisy}, a few initially cooperative players stay cooperative but they all die. These observations are all consistent with one of the underlying motivations for quorum sensing; bacteria waste energy if they engage in expensive cooperative behaviour when the population is too sparse.

\begin{figure}[!t]
	\centering
	\includegraphics[width=3.5in]{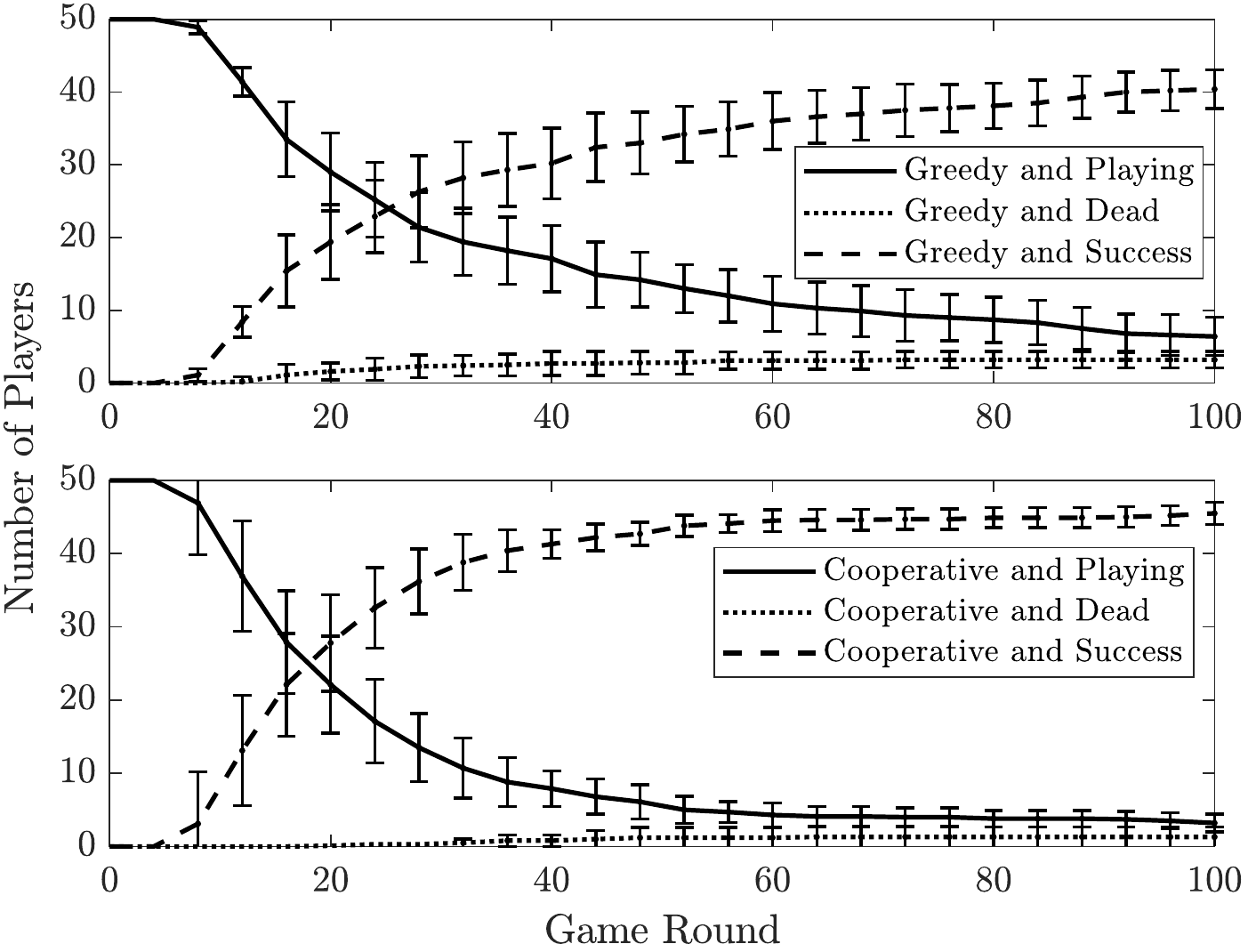}
	\caption{\textbf{Behaviour with high density and no switching.} Most players eventually succeed, albeit slowly, and some players (both greedy and cooperative) die.}
	\label{fig_bacteria_game_high_den_high_switch}
\end{figure}

\begin{figure}[!t]
	\centering
	\includegraphics[width=3.5in]{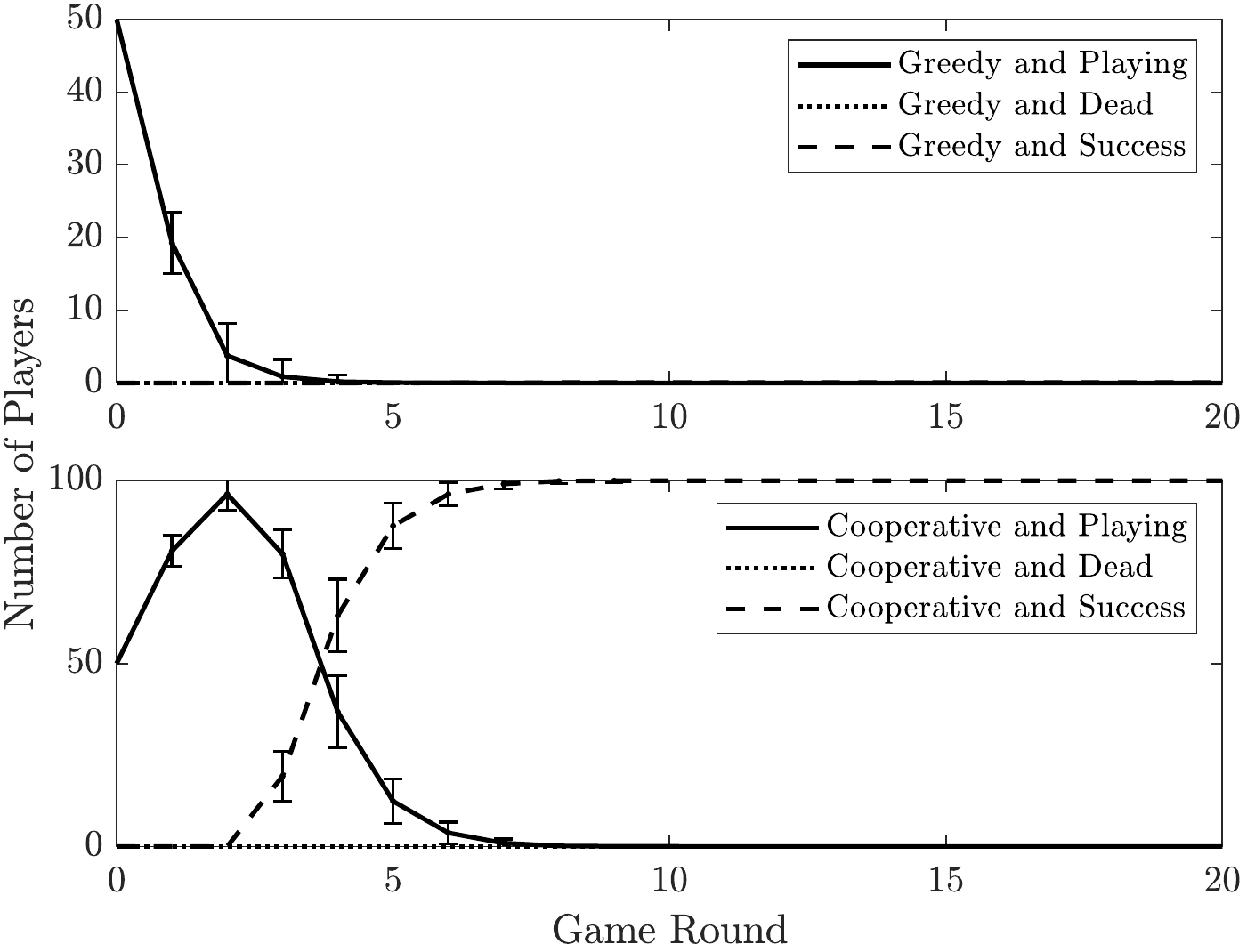}
	\caption{\textbf{Game with high density.} All players cooperate and succeed.}
	\label{fig_bacteria_game_high_den_low_switch}
\end{figure}

\begin{figure}[!t]
	\centering
	\includegraphics[width=3.5in]{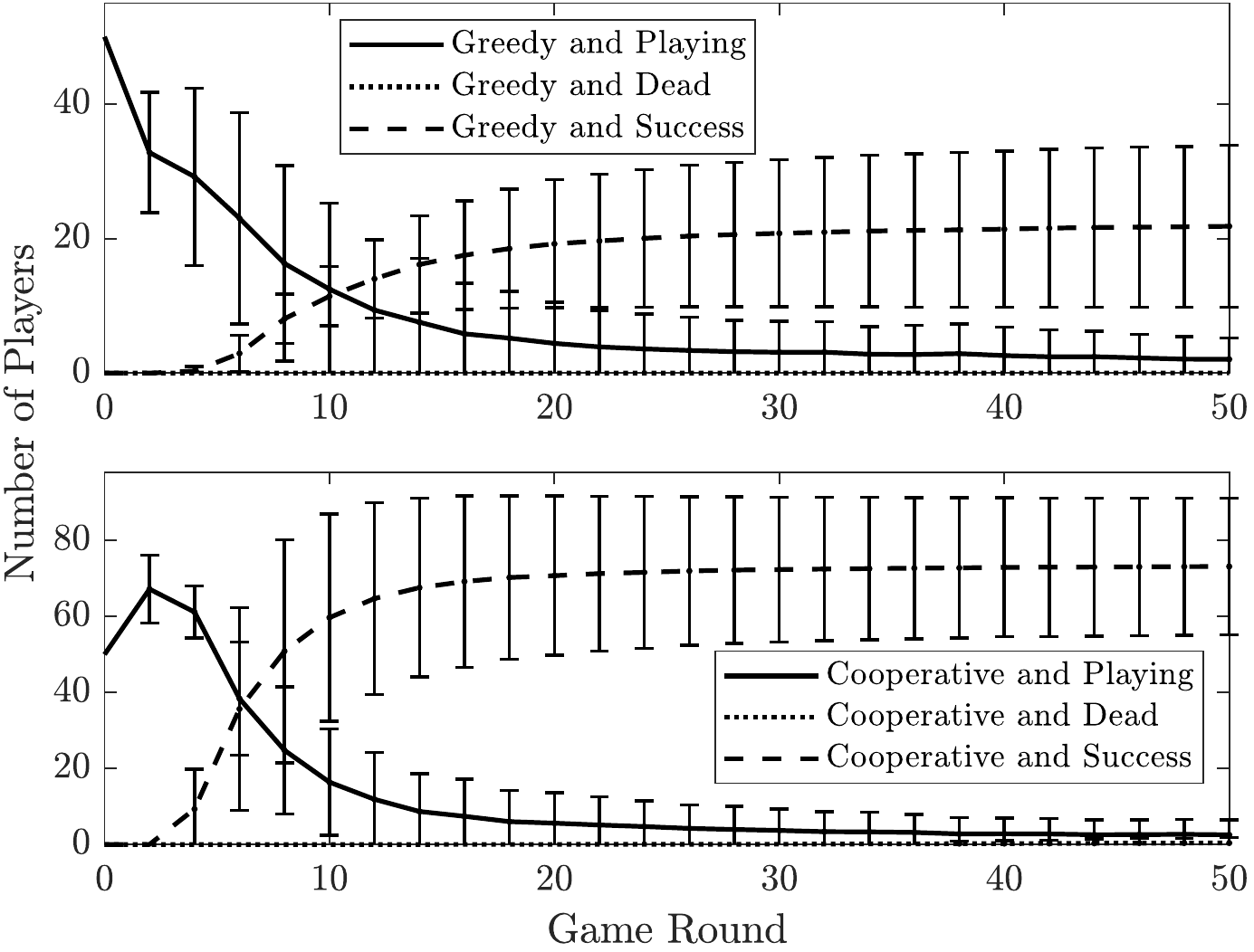}
	\caption{\textbf{Game with high density and estimation with AWGN.} Most players eventually succeed, including players that remained greedy (about 20\% of the total population).}
	\label{fig_bacteria_game_high_den_low_switch_noisy}
\end{figure}

We consider a high average population density in Figs.~\ref{fig_bacteria_game_high_den_high_switch}, \ref{fig_bacteria_game_high_den_low_switch}, and \ref{fig_bacteria_game_high_den_low_switch_noisy}, where the players are distributed over a square of length $10\,\mu$m. Based on our model, we expect to see successful cooperators. This is confirmed in Fig.~\ref{fig_bacteria_game_high_den_high_switch}, where the players are prevented from switching their behaviours. Nevertheless, the dynamics are more interesting than in the low density case. A clear majority of cooperative players become successful, but they are also able to support a majority of the greedy players to also succeed. Interestingly, a small fraction of both subpopulations dies. This can occur in a local area where a cooperator is isolated or where too many greedy players are close together.

When the switching costs are lowered in the high population density case, as considered in Fig.~\ref{fig_bacteria_game_high_den_low_switch}, all of the players cooperate and quickly succeed. We also observe this for \emph{any} initial behaviour distribution (i.e., players switch to cooperate even when they are all greedy; not shown). However, when we then introduce AWGN to the population estimation in Fig.~\ref{fig_bacteria_game_high_den_low_switch_noisy}, most players succeed within 50 rounds but a significant fraction (on average about 20\%) remain greedy. From our simple case study model, we see that noisy signalling information is sufficient to both \emph{create} and \emph{maintain} heterogeneous behavioural dynamics.

\begin{figure}[!t]
	\centering
	\includegraphics[width=3.5in]{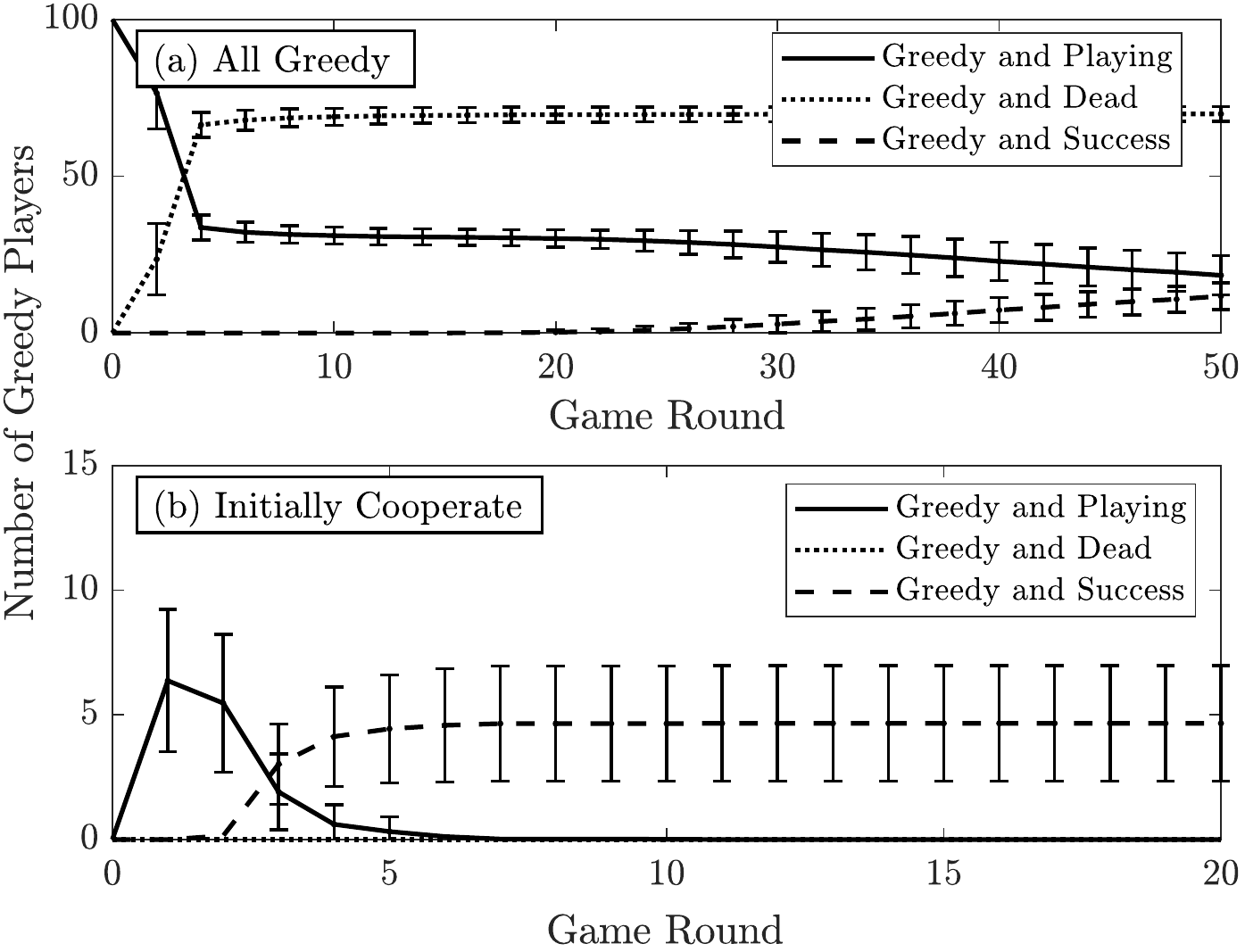}
	\caption{\textbf{Games with uniform initial behaviour.} High density environment where (a) all players are forcibly greedy; (b) all players initially cooperate but estimate density with AWGN.}
	\label{fig_bacteria_game_high_dens_uni_behaviour}
\end{figure}

Finally, in Fig.~\ref{fig_bacteria_game_high_dens_uni_behaviour} we test two of the questions raised by the high density results. First, are cooperators essential for the survival of greedy players? In Fig.~\ref{fig_bacteria_game_high_dens_uni_behaviour}(a), we initialize the players to be greedy and prevent switching. Most players quickly die and some that survive eventually succeed. Thus, the greedy players in a dense environment need cooperative players so they can free-ride. Second, can greed still emerge in a population that is initially fully cooperative? In Fig.~\ref{fig_bacteria_game_high_dens_uni_behaviour}(b), we observe the greedy behaviour when the population is initialized with cooperative players that make population estimates with AWGN. About 5\% of the population becomes greedy due to the noisy estimation, and as in Figs.~\ref{fig_bacteria_game_high_den_high_switch} and \ref{fig_bacteria_game_high_den_low_switch_noisy} they are able to succeed by free-riding off of the cooperators.

\subsection{Future Directions}

This simple case study has many opportunities for further extension. We suggest some of these here, any of which could be integrated within the current framework. In particular, we could model molecule propagation with more precision, i.e., by explicitly including the stochastic reaction-diffusion dynamics that accompany low local molecule concentrations. Other uncertain quantities for individual players might include the environment resource level or the local quantity of stored energy. In terms of the system model, we could consider environments with heterogeneous types of players, each with their own selection of strategies. More complex strategies might include player mobility (e.g., to move towards a food source), additional types of signals (e.g., for parallel estimation and coordination), and reproduction.

In terms of the analysis, we could consider the impact of different initial player distributions or the size of the population. In particular, we are interested in scaling up the populations by at least a few orders of magnitude. This section focused on monitoring the behaviour distribution as games progressed; we could alternatively consider the system equilibrium or the time until equilibrium as a function of the system parameters, such as the costs. We could also monitor the specific locations of the players in addition to their individual behaviour. Last but not least, we seek optimization problems and analytical results, including game theoretical analysis, to support the simulations and provide additional insight into the control of microscopic populations.

\section{Conclusions}
\label{sec_conclusions}

In this paper we identified opportunities to integrate game theoretic modelling with noisy signalling for real-time behavioural dynamics in microscopic environments. Integrating game theory and molecular communication can help us understand and possibly manipulate the competitive dynamics at a physical scale that accounts for the actions taken by individual cells. We identified how game theoretic models based on our approach are distinct from existing microscopic games in that they account for both \emph{real-time} and \emph{local} behaviour with \emph{noisy} information. We presented bacteria resource sharing and tumour cell signalling as two sample applications whose analysis and understanding could benefit from this integrated approach and potentially lead to control. In particular, we used a simple bacteria resource sharing game as a case study with an analytical model and simulations. We anticipate that many other microscopic scenarios could also benefit.

\section{Acknowledgment}

This work was supported in part by the Natural Sciences and Engineering Research Council of Canada (NSERC) through a postdoctoral fellowship.

\appendix

\section{Case Study System Model and Game Formulation}

In this Appendix, we present the detailed system model of the bacteria signalling case study in Section~\ref{sec_case_study} and describe how a game proceeds. We also list all of the parameter values used in the simulations in Table~\ref{table_sim_parameters}.

\begin{table}[!t]
	\centering
	\caption{Simulation Parameters Used in Section~\ref{sec_case_study}. Energy and cost parameters have arbitrary units.}
	
	{\renewcommand{\arraystretch}{1.4}
		\begin{tabular}{l||p{1.5cm}|p{1.8cm}}
			\hline
			Parameter & Symbol & Value \\ \hline \hline
			\# of Realizations per Game & -- & 100 \\ \hline
			Population Size & $\numPop$ & 100 \\ \hline
			Resource Strength & $\resource$ & 1\\ \hline
			Initial Energy & $\energy{0}$ & 50\\ \hline
			Starving Energy & $\energy{\mathrm{l}}$ & 0\\ \hline
			Success Energy & $\energy{\mathrm{u}}$ & 100\\ \hline
			Operating Cost & $\{\costCoop,\costGreedy\}$ & $\{5,1\}$\\ \hline
			Individual Conversion Factor & $\{\selfConversion{\coop},\selfConversion{\greedy}\}$ & $\{1,10\}$\\ \hline
			Inter-Player Conversion Factor & $\{\interCoop{\coop,\coop},\interCoop{\coop,\greedy},$ $ \interCoop{\greedy,\coop},\interCoop{\greedy,\greedy}\}$ & $\{10,5,-5,$ $-10\} \mu\mathrm{m}^{2}$\\ \hline
			Nominal Minimum Distance & $\distMin$ & $2\,\mu$m\\ \hline
			Behaviour Switching Cost & $\{\costSwitch{\coop},\costSwitch{\greedy}\}$ & $\{20,10\}$
		\end{tabular}
	}
	\label{table_sim_parameters}
\end{table}

\subsection{System Model}

A population $\pop$ of $\numPop$ bacteria is uniformly distributed over an environment. The environment provides an ambient resource strength $\resource$. The $j$th bacterium is a player with energy $\energy{j}$ that will starve if its energy reduces to $\energy{\mathrm{l}}$ and succeed if its energy increases to $\energy{\mathrm{u}}$.

Time progresses through a series of rounds of arbitrary length. In practice, a round should be on the order of the time required to modify gene expression (i.e., minutes or hours; see \cite{Alberts2015}), and much longer than the typical time for signalling molecules to diffuse across the population (i.e., seconds for a microscale population). In the $m$th round, the energy is updated as
\begin{equation}
\energy{j}[m] = \energy{j}[m-1] + \reward{j}[m] - \cost{},
\label{eqn_energy_update}
\end{equation}
where $\energy{j}[0] = \energy{0}$, $\reward{j}[m] = f(\resource)$ is the reward to the $j$th player, and $\cost{}$ is the cost in energy for the player to maintain its current behaviour. A player can be cooperative or greedy. Cooperation is more expensive than greed, i.e., $\costCoop > \costGreedy$. We scale the reward $\reward{j}$ by a cooperation scaling factor $\coopFactor{j}$, such that
\begin{equation}
\reward{j}[m] = \coopFactor{j}[m]\resource,
\end{equation}
where $\coopFactor{j}$ depends on the behaviour of all of the players. Cell fitness has been shown to depend on the proximity of other cells; see \cite{Lindsay2018}. A simple model that accounts for this is
\begin{equation}
\coopFactor{j}[m] = \selfConversion{j} + \sum_{k \in \pop / j}\frac{\interCoop{k,j}}{(\dist{k,j} + \distMin)^2},
\label{eqn_cooperative}
\end{equation}
where a player's individual capacity ($\selfConversion{j}$) to convert the resource to energy is added to the impact of all of the other players in the population $\pop$. The individual conversion capacity can take values $\selfConversion{j} \in \{\selfConversion{\coop},\selfConversion{\greedy}\}$, depending on whether the player is cooperative or greedy, respectively. $\interCoop{k,j}$ is a measure of how the $k$th player's behaviour affects the energy conversion of the $j$th player. $\dist{k,j}$ is the distance between the $j$th and $k$th players, and $\distMin$ is a nominal minimum distance that we choose to be about the size of a player. Thus, the impact of another player is inversely proportional to the square of the distance to that player. We choose this because the propagation time of a diffusing molecule increases with the square of the distance; see \cite{Berg1993}.

The inter-player conversion $\interCoop{k,j}$ can take values $\interCoop{a,b}$, where $a \in \{\coop,\greedy\}$ is the behaviour of the $k$th player and $b \in \{\coop,\greedy\}$ is the behaviour of the $j$th player. In this case study, we choose the cooperation scaling factor $\coopFactor{j}$ so that cooperative players benefit by being close to each other whereas greedy players benefit by being separated.

Thus, to achieve the desired impact of proximity on the cooperation scaling factor in (\ref{eqn_cooperative}), we impose that $\interCoop{\coop,\coop} > 0$, $\interCoop{\coop,\greedy} > 0$, $\interCoop{\greedy,\greedy} < 0$, and $\interCoop{\greedy,\coop} < 0$, i.e., a cooperative player always improves the conversion of nearby players and a greedy player always degrades the conversion of nearby players. The relative values are tunable, but in this work we consider that $\interCoop{\coop,\coop} > \interCoop{\coop,\greedy}$ (i.e., a player will receive a greater reward due to a cooperator if it also cooperates) and that $\interCoop{\greedy,\greedy} < \interCoop{\greedy,\coop}$ (i.e., a cooperating player is more resilient to the presence of a greedy player). Furthermore, we assume that a starved player (energy below $\energy{\mathrm{l}}$) is dead and no longer has an impact on the energy conversion of the other players, whereas a successful player (energy above $\energy{\mathrm{u}}$) continues the behaviour that made it successful, such that it affects other players as if it were still playing.

While (\ref{eqn_cooperative}) may be a reasonable approximation for how bacterial players affect each other, it would be unreasonable to expect that the players could accurately evaluate (\ref{eqn_cooperative}) with perfect information about the environment. In particular, a bacterium would not have perfect knowledge of the location and behaviour of all other bacteria in the population, although it may be able to infer some information from the propagation of signalling molecules; see \cite{Ruiz2018}. To keep things simple in this case study, we assume that each bacterium makes a noisy estimate of $\coopFactor{j}[m]$, $\coopFactorEst{j}[m]$, that is impaired by zero-mean Additive White Gaussian Noise with variance equal to $\coopFactor{j}[m]$. From $\coopFactorEst{j}[m]$, the player estimates the population size by assuming that the behaviour of the rest of the population is \emph{homogeneous} and that all players are \emph{adjacent}, i.e., $\dist{k,j} = 0$. From (\ref{eqn_cooperative}), the population estimate $\numPopEst{j}[m]$ is then
\begin{equation}
\numPopEst{j}[m] = \frac{(\coopFactorEst{j}[m] - \selfConversion{j})\distMin^2}{\interCoop{k,j}},
\end{equation}
where the inter-player conversion $\interCoop{k,j}$ depends on the player's current behaviour and the sign of $(\coopFactorEst{j}[m] - \selfConversion{j})$, i.e., if $(\coopFactorEst{j}[m] - \selfConversion{j})$ is positive, then the player assumes that the rest of the population is cooperative.

\subsection{Game Formulation}

The proposed game proceeds as follows. In every round, each player makes it own estimate $\numPopEst{j}[m]$ of the population size. It then decides whether to switch from its current behaviour by comparing the payoff in (\ref{eqn_energy_update}) as a greedy player and that as a cooperative player, \emph{assuming that all other (estimated) players would follow the same behaviour}. This comparison includes the cost $\cost{}$ associated with each behaviour. If a player decides to switch behaviours in order to increase its payoff, then it also has to pay a \emph{switching cost}. $\costSwitch{\coop}$ is the cost to switch from greedy to cooperative, and $\costSwitch{\greedy}$ is the cost to switch from cooperative to greedy. We do not include the switching cost when determining the preferred behaviour, such that players try to optimize their long-term behaviour, but we prevent a switch if the player would choose to switch but has insufficient energy to do so (i.e., if the switch would bring its energy level below $\energy{\mathrm{l}}$). The rounds continue until all the players are starved or successful, or until a maximum number of rounds has occurred.

\bibliography{gt_refs}

\end{document}